\documentclass[a4paper]{panl}
\usepackage{cite}
\usepackage{wrapfig}
\usepackage{graphicx}
\usepackage{amssymb}
\usepackage{amsfonts}
\usepackage{amsmath}
\usepackage{longtable}
\usepackage{rotating}
\usepackage{lscape}
\usepackage{epsfig}
\usepackage{multirow}
\usepackage[hypcap=false]{caption}
\originalTeX
\newcommand{\beq}[1]{\begin{equation} \label{#1}}
\newcommand{\eeq}   {\end{equation}}

\newcommand{\Frac}[2]{\frac
{\textstyle\lefteqn{\phantom{{}_{\mathstrut}^{\mathstrut}}} #1}
{\textstyle\lefteqn{\phantom{{}_{\mathstrut}^{\mathstrut}}} #2}}

\newcommand{\picref}[1]{fig.~\ref{#1}}
\graphicspath{ {./img/} }

\begin{document}

\title{Fast way to determine pp-collision time at the SPD experiment}
\maketitle
\authors{P. G. \,Filonchik $^{a,b,}$\footnote{E-mail: filonchik.pg@phystech.edu},
M. V.\,Zhabitsky $^{b,}$\footnote{E-mail: mikhail.zhabitsky@jinr.ru}}
\from{$^{a}$\,Moscow Institute of Physics and Technology, Moscow}
\from{$^{b}$\,Joint Institute for Nuclear Research, Dubna}

\begin{abstract}
Основная цель данной работы~--- найти быстрый и надёжный способ определения времени столкновения протонов в эксперименте SPD.
Основываясь на физике процесса, из входного потока реконструированных треков частиц мы производим выборку пионов, которая используется для вычисления несмещённой оценки времени столкновения. 
Точность оценки составляет около 30~пс. 
Метод является быстрым (не более 300~нс на одно событие) и надёжным, что позволит обрабатывать большой поток входных событий в эксперименте SPD. \
\vspace{0.2cm}

The main task of this work is to find a fast and robust way to determine $pp$--collision time $t_0$ at the SPD experiment. 
Using physics motivations, from the input flux of reconstructed particles' tracks we identify a subset of pions which is used to calculate the unbiased estimation of the event collision time. The uncertainty of the estimation is about 30~ps. This method is fast (less than 300~ns per event) and reliable, thus it will allow to process the high flux of input events at the SPD experiment.
\end{abstract}
\vspace*{6pt}


\label{sec:intro}
\section*{Introduction}

The Spin Physics Detector, one of the two facilities of the future NICA collider at the Joint Institute for Nuclear Research, is for studying the nucleon spin structure and spin-related phenomena with polarized proton and deuteron beams \cite{cdr}.
Understanding how the dynamics of the quarks and gluons determine the structure and the fundamental properties of the nucleon is one of the interesting unsolved problems of QCD.

The main task of this work is to determine $pp$--collision time based on measurements by the Time-Of-Flight (TOF) detector. 
One can solve this problem by combining information about reconstructed tracks with corresponding time measurements by the TOF detector. 
The $pp$--collision time allows us to reconstruct tracks with high accuracy and perform particle identification.

The idea of this project is to find a fast simple method to obtain an unbiased estimation of the $pp$--collision time.
We incorporate a priori knowledge about the process to accelerate the solution of the problem.

\label{sec:determination}
\section*{Determination of {\it pp}--collision time}
The $pp$--collision time~$t_0$ can be reconstructed from the time measurements by the TOF detector.
In the SPD experiment, the TOF detector will have a cylindrical geometry: two end-caps and a barrel with a radius of 1~m and a length of about 3~m~\cite{cdr}.
The detector's geometry provides complete coverage except two circular regions around the beam pipe.
A short distance between the collision point and the TOF
dictates the requirement for the TOF time resolution to be better than $\sigma_{t}=70$~ps.
The $t_0$ can be determined for multitrack events, when several particles originated from the same $pp$-vertex intersect the active area of the TOF detector, thus the detector measures hit moments $t_i$.
For $pp$-collisions at $\sqrt{s}=27~\text{GeV}$ events of interest will have a multiplicity of more than 5~charged tracks. 
By the track reconstruction, the hit in the TOF is linked to the corresponding track, its length~$L_i$ and momentum~$p_i$ will be measured by the SPD Vertex and Straw detectors.
For charged particles with $p_\perp>0.5~\text{GeV}/c$ the relative precision of the momenta measurement is required to be 2\%. 

The expected time of flight (tof) for a particle with mass $m_i$ reads
\beq{tof}
    t_{\text{tof}}(m_i) = \frac{L_i}{c} \sqrt{1+\frac{m_i^2 c^2}{p^2}},
\eeq
where a mass hypothesis $m_i$ can correspond to a charged $\pi$, $K$ or proton.
At the SPD conditions, relativistic pions with momenta higher than $0.5~\text{GeV}/c$ can not be distinguished from electrons by their time of flight.
Other particle types are rare and corresponding mass hypotheses can be taken into account as will be shown later.

To find the $pp$--collision time $t_0$ in the event with $n$ reconstructed charged tracks we minimize the weighted sum of squared residuals between time measurements and the predicted times of particles crossings the TOF detector:
\beq{chi2}
\chi^2(\{m_i\}_n) = \sum_{i=1}^n \Frac{(t_0 + t_{\text{tof}}(m_i) - t_i)^2}
    {\sigma_t^2 + \sigma_{\text{tof}_i}^2} =
    \sum_{i=1}^n \Frac{(t_0 - t_{\text{diff}}(m_i))^2}
    {\sigma_t^2 + \sigma_{\text{tof}_i}^2}. 
\eeq
If all particle types were known the estimation of $pp$--collision time is
\beq{t0}
    \hat{t}_0 = \sum_i \frac{t_{\text{diff}}(m_i)}{\sigma_t^2 + \sigma_{\text{tof}_i}^2} \cdot \left(\sum_i \frac{1}{\sigma_t^2 + \sigma_{tof_i}^2} \right)^{-1}.
\eeq
Uncertainty $\sigma_{tof_i}$ in the predicted time of flight is mainly due to momentum resolution ($\sigma_p / p \sim 0.02$).
For fast particles $\sigma_{tof_i}$ is much smaller than the time resolution of the TOF detector ($\sigma_{t}=70$~ps).
The latter doesn't depend on the particle's type, thus a mean of $t_{\text{diff}}(m_i)$ can serve as a good estimate of the $pp$--collision time.

According to Eq.~(\ref{chi2}), the determination of $pp$--collision time is an optimization problem to minimize $\chi^2$ as a function of $t_0$ and mass hypotheses $\{m_i\}_n$ for reconstructed charged tracks in the event.
One can try to solve~(\ref{chi2}) by a brute-force search, which checks all possible combinations of particles and selects the one with the minimal $\chi^2$.
Even if only three types of particles are allowed ($\pi/K/p$), the brute-force algorithm has too high complexity of $O(3^n)$ operations, thus it takes up to few seconds to find the minimum for events with higher multiplicity.
To avoid an exhaustive search one can combine tracks in subgroups and find a near optimal solution faster, this approach was applied in works \cite{alice+time} and \cite{perfom+tof}.
Another direct approach is to incorporate some optimization algorithm, which will shuffle through the parameter space towards the minimum.
In the following, we report how a priori knowledge about the physics of $pp$-collisions can be used to accelerate the solution of the problem.




\label{sec:motivations}
\section*{Physics motivations and the "Sliding window"\ method}
\begin{figure}
\begin{center}
\includegraphics[width=127mm]{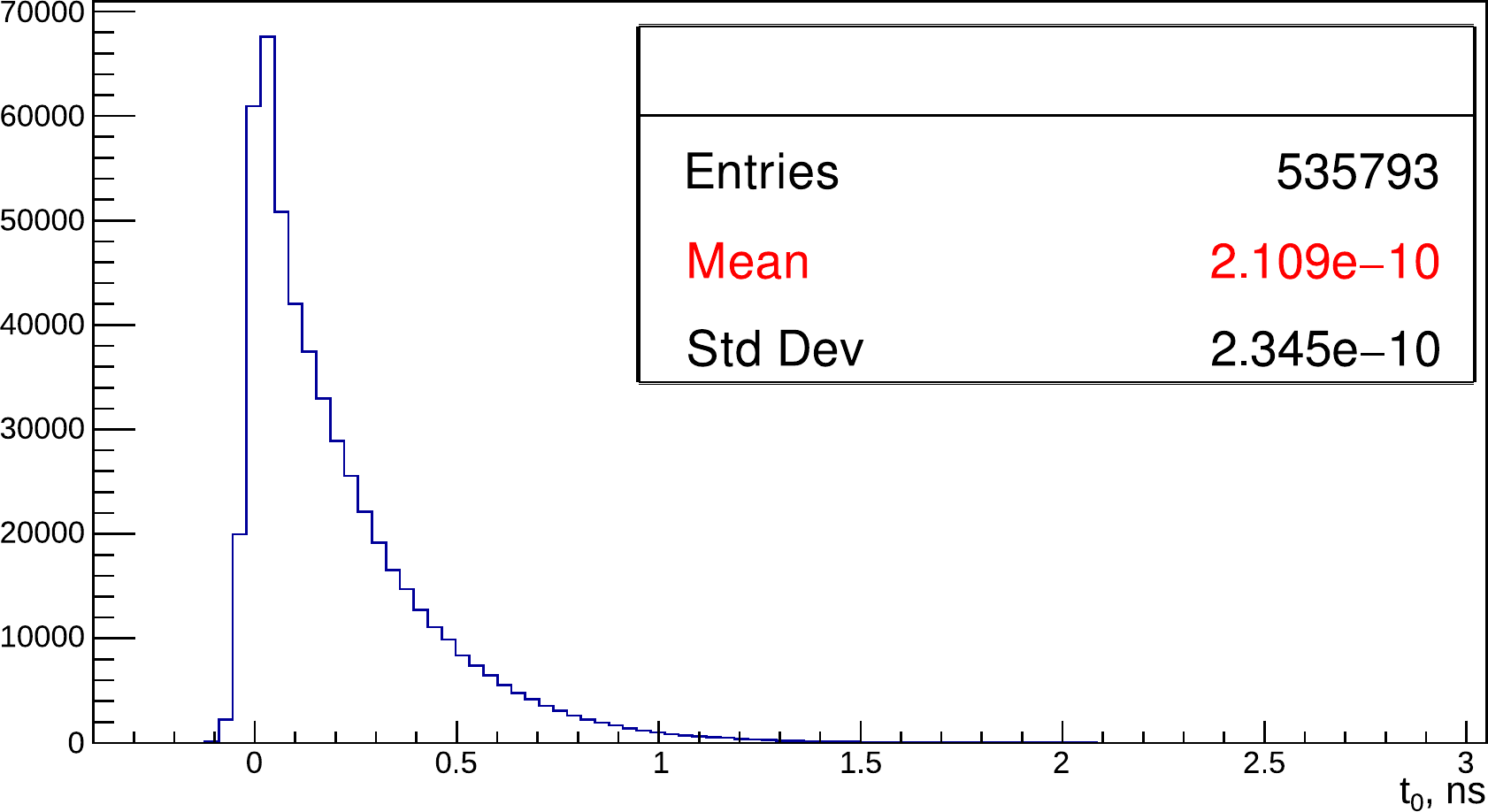}
\vspace{-3mm}
\caption{$t_0$-distribution under hypothesis that all particles are pions.}\label{t0 simple}
\end{center}
\vspace{-5mm}
\end{figure}

\begin{figure}
    \centering
\begin{minipage}[t]{0.46\textwidth}
   \centering
   \includegraphics[width=\textwidth]{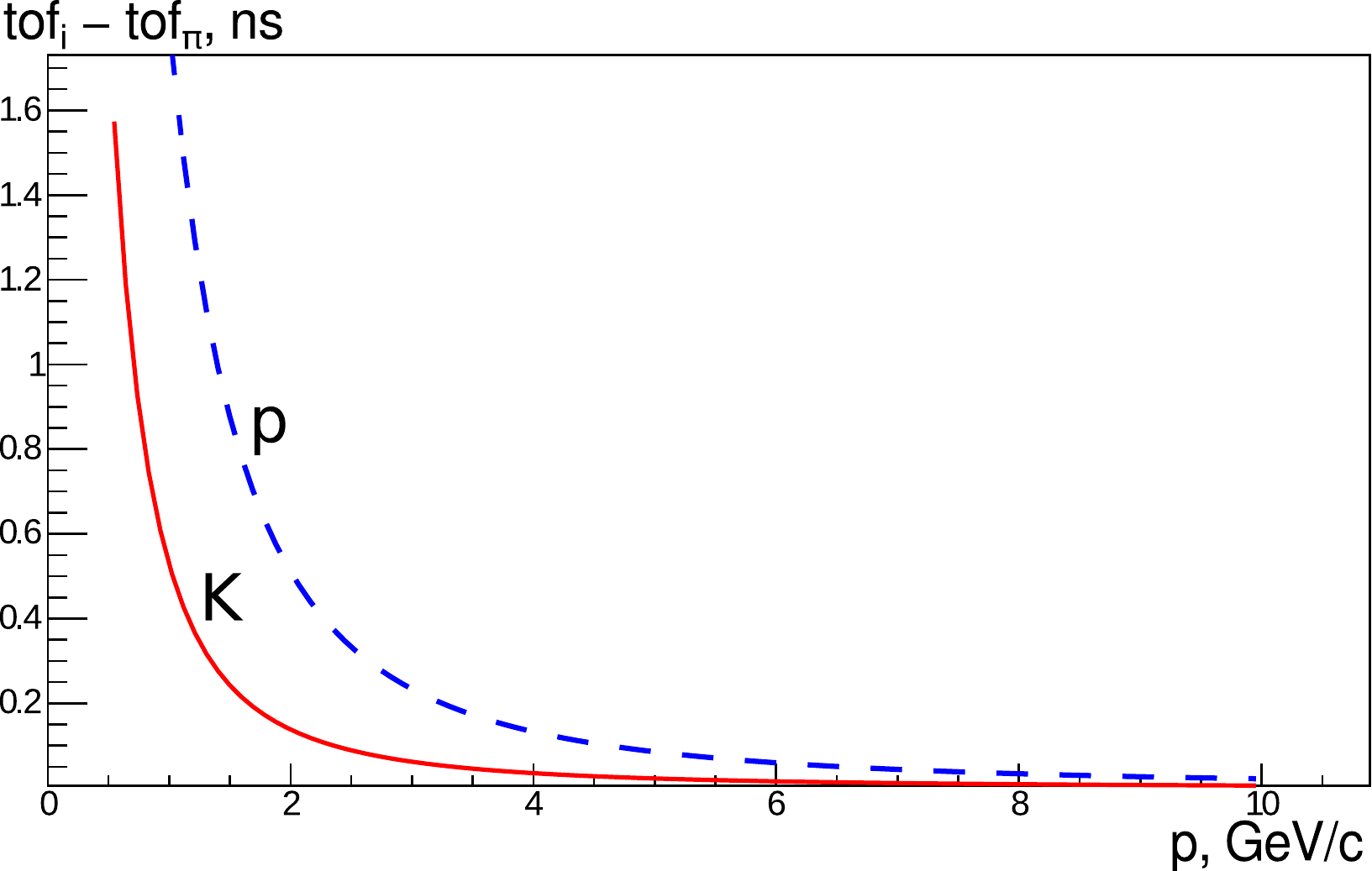}
   \vspace{-3mm}
   \captionof{figure}{Difference in TOF between kaons and pions (solid$~$line), protons and pions (dashed line).}
   \label{tdiff p k}
   \vspace{-5mm}
\end{minipage}
\hfill%
\begin{minipage}[t]{0.46\textwidth}
   \begin{center}
   \includegraphics[width=\textwidth]{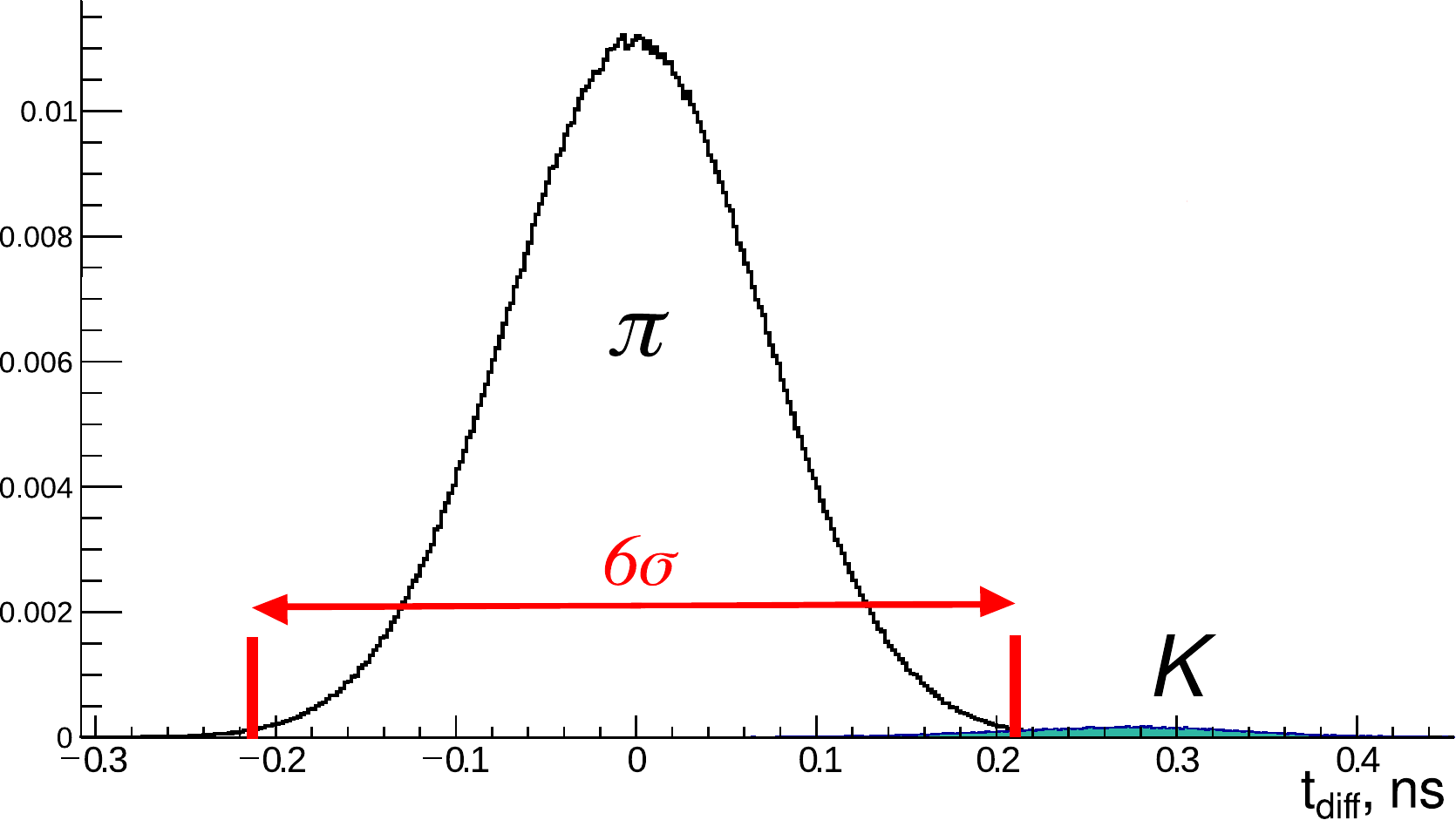}
   \vspace{-3mm}
  \captionof{figure}{Distributions of $t_{\text{diff}}$ for pions and misidentified kaons at momentum $1.5~\text{GeV}/c$.}
   \label{tdiff pi k}
   \vspace{-5mm}
   \end{center}
\end{minipage}
\vspace{1mm}
\end{figure}

Pions are the most abundant secondary particles originated from inelastic $pp$-collisions at $\sqrt{s}=27~\text{GeV}$.
Instead of checking all possible mass combinations in eq.~(\ref{chi2}), one can attribute the pion mass to every charged particle in order to estimate the time of $pp$-collision. 
Without surprise the distribution of reconstructed $\hat{t}_0$ exhibits a big tail in this case (\picref{t0 simple}).
As masses of misidentified kaons or protons are replaced by the smaller pion's mass, their estimated times of flight are shifted to shorter times, thus this simplification results in $\hat{t}_0$ estimations biased to delayed values.
The difference in time~of~flight between heavier particles and the pion is more pronounced at softer momenta, while relativistic particles can not be distinguished by their time of flights at short distances between the collision point and the TOF detector at the SPD (\picref{tdiff p k}).

At momentum $1.5~\text{GeV}/c$ kaon's time of flight is by about $0.2$~ns longer than for the pion with the same momentum.
Even slower kaons are delayed by more than $3\sigma_t$ with respect to pions (\picref{tdiff pi k}).
If one sorts all reconstructed particles with momenta below $p_{\text{max}} = 1.5~\text{GeV}/c$ by their $t_{\text{diff}}$ under pion's mass hypothesis, pions will be at the lower range and almost all of them fall into $\pm 3\sigma_t$ range around unknown $t_0$.
At the same time misidentified heavy particles will be delayed by at least $3\sigma_t$ and scattered.
As the next step one can slide a window of the $6\sigma_t$ length along the sorted $t_{\text{diff}}$ in order to identify a window's position with the most $t_{\text{diff}}$ inside it.
As pions are by far the most abundant secondaries, in most events the tracks inside the found range will represent a pure pions sample.
Then the $pp$--collision time is estimated as a mean of timings, which have fallen into the $6\sigma_t$ window.
The method requires at least 3~tracks to be within the search window,
about 90\% of events fulfill this criterion.  



\begin{figure}[h]
   \begin{center}
   \includegraphics[width=127mm]{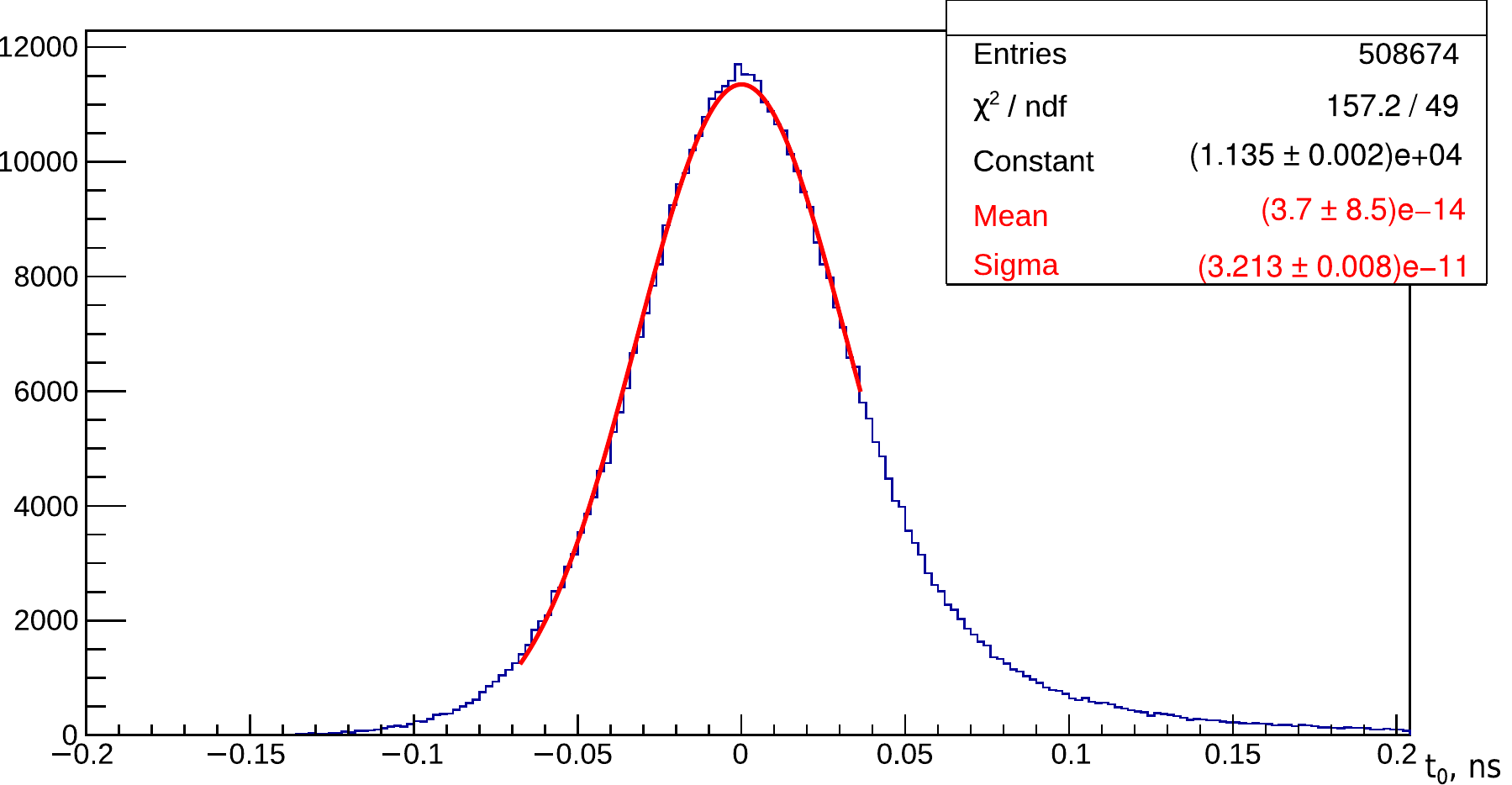}
   \vspace{-3mm}
   \caption{Reconstructed $\hat{t}_0$-distribution by the sliding window method}
   \label{t0 best}
   \end{center}
   \vspace{-5mm}
\end{figure}

The distribution of $pp$--collision times $\hat{t}_0$ obtained by the sliding window method is shown in \picref{t0 best}. 
The $\hat{t}_0$-estimation is unbiased with a resolution of about $32$~ps.
The sliding window method provides $pp$-collision time for track reconstruction and for particle identification by TOF.
The typical programme execution time is about 300~ns.
Thus the sliding window method allows to process a high flux of input events at the SPD.

The momentum range to select reconstructed tracks was chosen in the way, to ensure the estimation of event collision time $\hat{t}_0$ be unbiased. As the cross section of pions inclusively produced in $pp$-collisions is peaked at momenta below $1~\text{GeV}/c$, an increase of the upper momentum limit $p_{\text{max}}$ will only slightly improve the statistical uncertainty of $\hat{t}_0$ estimations, but will introduce a bias towards delayed $\hat{t}_0$ due to misidentified kaons. 

\section*{Results and conclusions}
In this work, we propose the sliding window method to reconstruct the time of $pp$-collisions from the measurements by the SPD TOF detector.
From the input flux of reconstructed tracks, we identify a pure sample of pions which is used to calculate the unbiased estimation of the event collision time with a resolution of about $32$~ns.
The sliding window method can provide $pp$-collision time for track reconstruction and for particle identification by their time of flight.
The typical time to find $t_0$ by the sliding window method is about 300~ns.
The fast determination of $pp$-collision time allows to process the high flux of input events at the SPD experiment.

This work was supported by the JINR START programme.
The authors would like to thank Semyon Yurchenko and other members of the SPD Collaboration for fruitful discussions and support.


\bibliographystyle{pepan}
\bibliography{pepan_biblio}

\begin{thebibliography}{1}
\def\selectlanguageifdefined#1{
\expandafter\ifx\csname date#1\endcsname\relax
\else\selectlanguage{#1}\fi}
\providecommand*{\href}[2]{{\small #2}}
\providecommand*{\url}[1]{{\small #1}}
\providecommand*{\BibUrl}[1]{\url{#1}}
\providecommand{\BibAnnote}[1]{}
\providecommand*{\BibEmph}[1]{\emph{#1}}
\ProvideTextCommandDefault{\cyrdash}{\hbox to.8em{--\hss--}}
\providecommand*{\BibDash}{\ifdim\lastskip>0pt\unskip\nobreak\hskip.2em\fi
\cyrdash\hskip.2em\ignorespaces}

\bibitem{cdr}
\selectlanguageifdefined{english}
\BibEmph{Abazov V.M. et~al.} [SPD Collaboration] Conceptual design of the Spin
  Physics Detector. \BibDash
\newblock 2021. \BibDash
\newblock URL: \BibUrl{https://arxiv.org/abs/2102.00442}.

\bibitem{alice+time}
\selectlanguageifdefined{english}
\BibEmph{{ALICE collaboration}.} Determination of the event collision time with
  the ALICE Detector at the LHC~//
  \href{http://dx.doi.org/10.1140/epjp/i2017-11279-1}{Eur. Phys. J. Plus}.
  \BibDash
\newblock 2017. \BibDash
\newblock V. 132. \BibDash
\newblock P.~1--17. \BibDash
\newblock 99.

\bibitem{perfom+tof}
\selectlanguageifdefined{english}
\BibEmph{Akindinov A., Andrea A., Agostinelli A., Antonioli P., Arcelli S.,
  Basile M., Bellini F., Romeo G., Cifarelli L., Cindolo F., Colocci M., Caro
  A., De~Gruttola D., De~Pasquale S., Doroud K., Fusco~Girard M., Guerzoni B.,
  Hatzifotiadou D., Kim D., Zichichi A.} Performance of the$~$ALICE
  Time-Of-Flight detector at the LHC~//
  \href{http://dx.doi.org/10.1140/epjp/i2013-13044-x}{Eur. Phys. J. Plus}.
  \BibDash
\newblock 2013. \BibDash
\newblock V. 128. \BibDash
\newblock P.~1--9. \BibDash
\newblock 44.

\end{thebibliography}

\end{document}